\documentclass[10pt, twocolumn,
superscriptaddress,
english,
prb,
showpacs,
floatfix,
aps]{revtex4-2}
\usepackage[backref=none,bookmarksnumbered=true,bookmarks=true,  	bookmarksopen=true,colorlinks=true,citecolor=blue,linkcolor=blue, 	anchorcolor=green,urlcolor=blue,unicode=false]{hyperref}

\usepackage{graphicx}
\usepackage{float}
\usepackage{dcolumn}
\usepackage{bm}
\usepackage{hyperref}
\usepackage{chemformula}
\usepackage{siunitx}
\usepackage{verbatim}
\usepackage[T1]{fontenc}
\usepackage[utf8]{inputenc}

\usepackage{times}
\usepackage[symbol]{footmisc}

\begin{document}

\author{Vibhuti N. Rai}
\email{vibhuti.rai@fu-berlin.de}
\affiliation{Institute for Quantum Materials and Technologies, Karlsruhe Institute of Technology (KIT), D-76344 Eggenstein-Leopoldshafen, Germany}
\author{Christof Holzer}
\affiliation{Institute for Quantum Materials and Technologies, Karlsruhe Institute of Technology (KIT), D-76344 Eggenstein-Leopoldshafen, Germany}
\affiliation{Institute of Theoretical Solid State Physics, Karlsruhe Institute of Technology (KIT), D-76131 Karlsruhe, Germany}
\author{Carsten Rockstuhl}
\affiliation{Institute of Theoretical Solid State Physics, Karlsruhe Institute of Technology (KIT), D-76131 Karlsruhe, Germany}
\affiliation{Institute of Nanotechnology, Karlsruhe Institute of Technology (KIT), D-76344 Eggenstein-Leopoldshafen, Germany}
\author{Wulf Wulfhekel}
\affiliation{Institute for Quantum Materials and Technologies, Karlsruhe Institute of Technology (KIT), D-76344 Eggenstein-Leopoldshafen, Germany}
\author{Lukas Gerhard}
\affiliation{Institute for Quantum Materials and Technologies, Karlsruhe Institute of Technology (KIT), D-76344 Eggenstein-Leopoldshafen, Germany}

\title{Gating upconversion electroluminescence in a single molecule via adsorption-induced interaction of unpaired spin}

\begin{abstract}
Molecules with unpaired spins (radicals) offer promising alternatives to closed-shell molecules as they are less limited regarding the spin statistics in their electroluminescence. Here, we combine scanning tunneling microscopy induced luminescence and density functional theory to study single vanadyl phthalocyanine molecules, which are stable neutral radicals. Two distinct adsorption geometries of the molecule on NaCl/Au(111) lead to a difference in the interaction of the unpaired electron with the substrate, which in turn allows us to investigate its effects on the light emission process. Remarkably, we observe that up-conversion electroluminescence is gated by the adsorption geometry of the molecule, an effect we attribute to a reordering of excited states and enhanced excited state transition probabilities. The profound influence of the unpaired electron via state reordering opens new possibilities for tuning not only molecular electroluminescence but also many other spin dependent phenomena.

\end{abstract}

\maketitle
\newpage

Understanding the process that converts an electrical current to light on the scale of a single molecule is necessary to exploit the full potential of molecular materials in light emitting applications. 
Most molecules tend to have a closed-shell electron configuration and a singlet spin configuration in their ground state. For such molecules, excitation via subsequent charge injections (CI) leads to 75$\,\%$ triplet excitons, which cannot recombine via photon emission without spin-orbit interaction, and only 25$\,\%$ singlet excitons contribute to the photon emission. This is one aspect why the class of molecules with an unpaired electron and a doublet spin configuration are particularly interesting to be exploited for luminescence, since in such radicals, all electrically accessible excited states are, in principle, spin-allowed to decay radiatively. 

In recent years, scanning tunneling microscope-induced luminescence (STML) has been shown to provide a controlled environment to explore such electroluminescence processes at the level of individual molecules \cite{qiu_vibrationally_2003,zhang_visualizing_2016,imada_real-space_2016,doppagne_vibronic_2017,zhang_sub-nanometre_2017, doppagne_electrofluorochromism_2018, kaiser_single-molecule_2019, dolezal_exciton-trion_2021,doppagne_single-molecule_2020,dolezal_evidence_2022,rai_boosting_2020,cao_energy_2021,rai_activating_2023,rai_hot_2023,jiang_many-body_2023,jiang_topologically_2023,kaiser_gating_2024,kaiser_electrically_2025}.
The prospect of new pathways for electroluminescence has triggered attempts to create radical versions of phthalocyanine (Pc) molecules that are most intensely studied in STML. Dehydrogenation of Pc on decoupled surfaces \cite{vasilev_internal_2022} led to isoelectronic deprotonated compounds without significantly affecting the frontier orbitals located on the macrocycle that participate in the excitation-emission cycle. However, the critical role of the interaction between the unpaired spin and the substrate in higher-order processes such as trionic emission \cite{doppagne_electrofluorochromism_2018, dolezal_mechano-optical_2020, rai_boosting_2020, dolezal_exciton-trion_2021, rai_activating_2023} or up-conversion electroluminescence (UCEL) \cite{chen_spin-triplet-mediated_2019,rai_boosting_2020,rai_hot_2023,luo_anomalously_2024} has not been investigated so far.

To this end, we employ vanadyl phthalocyanine (VOPc) molecules adsorbed on NaCl/Au(111) for a combined STML and density functional theory (DFT) investigation. VOPc is a stable neutral radical in the ground state with an unpaired electron. In contrast to widely studied planar Pc molecules, VOPc adopts two distinct adsorption configurations: Either with the central oxygen atom oriented toward the substrate (O-down), or with it pointing away (O-up) \cite{eguchi_molecular_2013,niu_molecular_2014,zhang_single_2014,malavolti_tunable_2018, kaiser_single-molecule_2019,blowey_structure_2019}. 
This dual geometry provides a unique framework for understanding the role of the unpaired electron in the light emission process. As the unpaired spin is mostly located at the vanadyl \cite{malavolti_tunable_2018, noh_template-directed_2023}, its interaction with the environment is significantly different for the two configurations, while the frontier orbitals on the macrocycle experience almost the same environment. In agreement with previous photoluminescence experiments \cite{debnath_triggering_2022}, but in contrast to STML experiments on VOPc/NaCl/Ag(111) \cite{kaiser_single-molecule_2019}, our STML experiments and DFT calculations show two emission bands originating from the neutral molecule and the cation. Similarly to previous work \cite{atzori_room-temperature_2016,malavolti_tunable_2018}, the frontier orbitals of the neutral molecule localized on the Pc macrocycle are basically unaffected by the orientation of the VOPc. However, we observed intense UCEL for O-down configuration but not for O-up, which we relate to the role of the unpaired spin in the charged state. Based on our DFT calculations and STML experiments, we provide a many-body description that explains how flipping the molecule upside down gates the UCEL via reordering of triplet states, an increased lifetime of the excited states, and an increased probability for excited state transitions calculated via quadratic response time-dependent density functional theory (TD-DFT).

This study opens new routes for efficient UCEL by parking the energy in dark states of the neutral molecule and exploiting the molecule-substrate interaction rather than relying on the molecular design alone.

\section{Results and discussion}

\begin{figure}
\centering
\includegraphics{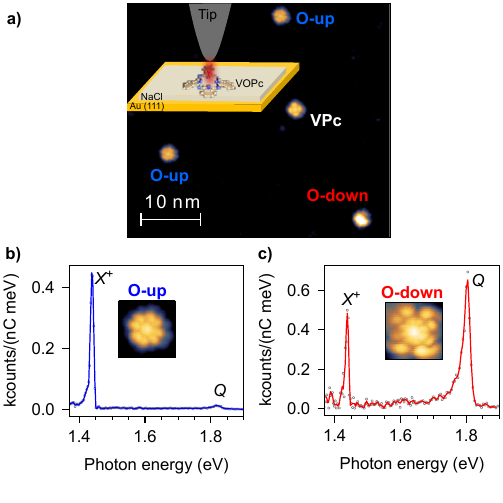}
\caption{\label{fig:Schematic_Topo} {\bf STML of VOPc in different adsorption configurations.}. (a) Schematic of the experimental setup (tip and molecule are not to scale) showing a single VOPc molecule decoupled from the Au(111) surface by 3ML of NaCl. (b) Topographic overview showing two distinct configurations of VOPc identified as O-up (blue) and O-down (red) ($V_\mathrm{b}$ = -2.0\,V, and $I$ = 1\,pA). (c) STML spectrum on an isolated O-up recorded with $V_\mathrm{b}$ = -2.20\,V, and $I$ = 65\,pA, $t_\mathrm{exp}$ = 100\,s. Emission lines at two distinct energies are labeled as $Q$ and $X^+$. Corresponding topography is shown in the inset ($V_\mathrm{b}$ = -2.0\,V, and $I$ = 1\,pA). (d) STML spectrum on an isolated O-down recorded with $V_\mathrm{b}$ = -2.05\,V, $I$ = 45.5\,pA, $t_\mathrm{exp}$ = 10\,s. Emission lines at two distinct energies are labeled as $Q$ and $X^+$. Corresponding topography is shown in the inset ($V_\mathrm{b}$ = -2.35\,V, and $I$ = 4\,pA).}
\end{figure}

Figure \ref{fig:Schematic_Topo}a illustrates the STML setup, with a VOPc molecule on 3 monolayers of NaCl on Au(111) (for details of the sample preparation and the light collection setup, see SI Note 1) \cite{edelmann_light_2018}. In agreement with previous work \cite{niu_molecular_2014,zhang_single_2014,blowey_structure_2019,kaiser_single-molecule_2019}, a typical topography reveals two distinct adsorption configurations of VOPc (see Fig.~\ref{fig:Schematic_Topo}b) that we tentatively assign as O-down/O-up, respectively, based on their higher/lower apparent height (also see insets of Fig.~\ref{fig:Schematic_Topo}c and d).
Although such assignment is not obvious based on the STM topography alone, this assumption provides a framework for understanding the observed STML from the two VOPc configurations, which will be further corroborated in this report.
As previously reported, we observed transitions between the two orientations induced by the STM tip \cite{zhang_single_2014}, confirming that these are chemically identical molecules.
Molecules without the central oxygen atom (VPc) can also be distinguished but are excluded from the following discussion as they show no STML \cite{kaiser_single-molecule_2019} (see Fig.~\ref{fig:Schematic_Topo}b and SI Note 1).

The two configurations with intact vanadyl (O-down/O-up) show STML with emission peaks at very similar energies when excited with high negative bias voltages $V_\mathrm{b}$ (see Fig.~\ref{fig:Schematic_Topo}c and d). The STML spectrum recorded on O-up shows two emission bands labeled $Q$ ($\hbar\omega \approx$ 1.80\,eV) and $X^+$ ($\hbar\omega \approx$ 1.45\,eV). $Q$ is known to be the photon emission of the charge-neutral state of the molecule \cite{kaiser_single-molecule_2019, debnath_triggering_2022}. $X^+$ can be identified as the emission of the positively charged state of the molecule ($\text{VOPc}^+$) based on previous charge-state-dependent photoluminescence measurements \cite{debnath_triggering_2022} and our DFT calculations, which will be discussed later in this report. 
The STML spectrum of O-down features the same two emission bands, $Q$ ($\approx$ 1.80\,eV) and $X^+$ ($\approx$ 1.45\,eV), as shown in Fig.~\ref{fig:Schematic_Topo}d. 
The emission energies of $X^+$ and $Q$ are very similar to those of closed-shell Pc molecules \cite{doppagne_electrofluorochromism_2018, rai_boosting_2020}.
The main $Q$ peak is significantly sharper in the O-down orientation than in the O-up orientation, about 5.8 meV compared to 40 meV, based on a Lorentzian fit (see Fig.~S1a-d). This suggests that the excited-state lifetime is significantly longer in O-down compared to O-up. The $X^+$ emission bands show similar peak widths in both configurations (limited by the detector, see SI Note 1). Higher resolution spectra also suggests that the emission bands have contributions from multiple peaks, which can be attributed to vibrons or libron modes (see Fig.~S1a) \cite{zhang_visualizing_2016,imada_real-space_2016,doppagne_electrofluorochromism_2018,doppagne_single-molecule_2020,rai_boosting_2020,kong_probing_2021,dolezal_evidence_2022,rai_hot_2023,friedrich_fluorescence_2024}.

\begin{figure*}
\centering
\includegraphics{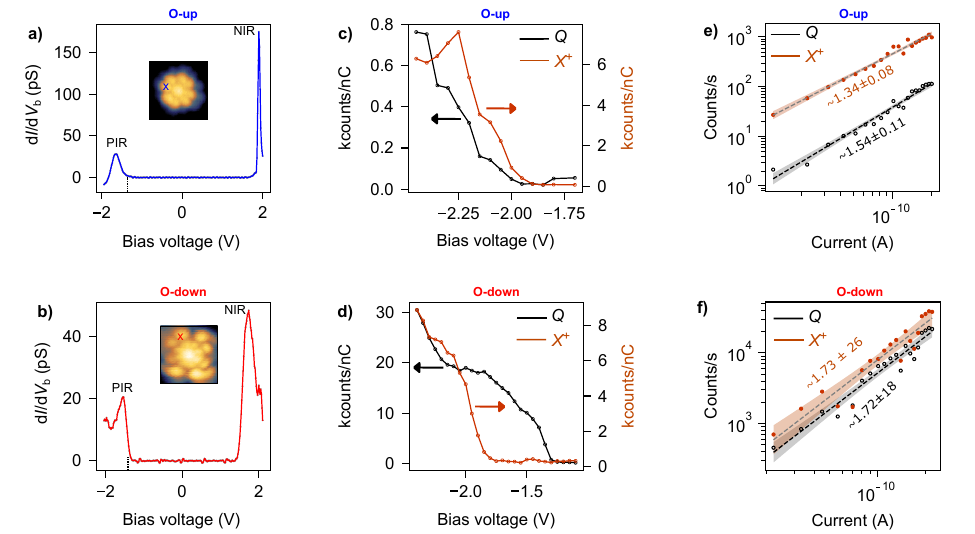}
\caption{\label{fig:Electronic_struc} {\bf Electronic structure of VOPc and energetics of the emission bands}. (a) and (b) $dI/dV_\mathrm{b}$ spectrum recorded on O-up and O-down, respectively (inset: tip positions are marked with a blue/red cross). Dotted lines indicate the onset of the PIR at -1.35\,V. (c) and (d) Integrated photon counts of $Q$ (black) and $X^+$ (brown) bands as a function of $V_\mathrm{b}$ recorded on O-up and O-down, respectively. Parameters for recording the spectra on O-up $I$ = 40\,pA, $t_\mathrm{exp}$ = 30\,s and on O-down $I$ = 45\,pA, $t_\mathrm{exp}$ = 10\,s. (e) and (f), Dependency of integrated photon counts of $Q_x$ (black) and $X^+$ (brown) emission bands on tunneling current at fixed bias voltages $V_\mathrm{b}$ = -2.20\,V (O-up) and $V_\mathrm{b}$ = -2.40\,V (O-down). Shaded areas show the 95\% confidence band for the fits. Parameters for recording the inset topographies in (a) and (b) are $V_\mathrm{b}$ = -2.0\,V, $I$ = 1\,pA and $V_\mathrm{b}$ = -2.35\,V, $I$ = 4\,pA, respectively.
} 
\end{figure*}

To understand the excitation process via CI, the electronic structure of VOPc plays a crucial role. Therefore, we performed differential conductance measurements ($dI/dV_\mathrm{b}$) on O-up and O-down configurations. In Fig.~~\ref{fig:Electronic_struc}a and b, the $dI/dV_\mathrm{b}$ spectra recorded on O-up and O-down VOPc show a peak at -1.65\,V and -1.52\,V, respectively, indicating positive ion resonance (PIR), both with an onset at about $V_\mathrm{b}$ $\approx$ -1.35\,V (also see Fig.~S2a). This means that transient positive charging becomes possible when $V_\mathrm{b}$ $\leq$ -1.35\,V, indicating that the positive charge is distributed over the macrocycle and barely affected by the orientations. 
For O-down, an additional peak emerges at $V_\mathrm{b}$ $\approx$ -2.0\,V. At positive voltages, O-up and O-down exhibit negative ion resonances (NIR) at 1.9\,V and 1.75\,V / 1.95\,V respectively, with an onset of 1.6\,V. The correlation between the emergence of STML and the energy-dependent $dI/dV_\mathrm{b}$ resonances forms the basis for the microscopic excitation mechanism discussed later. Note that the broadening of the ion resonances is not a reliable indicator for the lifetimes, but rather related to the coupling to the NaCl phonons\cite{repp_molecules_2005}.

Potential UCEL manifests as molecular emission of photons with energies $\hbar\omega$ higher than that of the exciting electrons, $eV_\mathrm{b}$.  
Therefore, we measured integrated photon counts of the exciton $Q$ ($\hbar\omega \approx$1.80 eV, black data points in Fig.~\ref{fig:Electronic_struc}c and d)) and trion $X^+$ ($\hbar\omega \approx$ 1.45\,eV, brown data points in Fig.~\ref{fig:Electronic_struc}c and d) emission bands as a function of $V_\mathrm{b}$. Molecules in O-up configuration show no light emission at -1.85\,V < $V_\mathrm{b}$ and at -2.2\,V < $V_\mathrm{b}$ < -1.85\,V, both $Q$ and $X^+$ lines become visible. Note that this threshold value for the onset of light emission lies significantly below the PIR onset of $\approx$ -1.35\,V, \textit{i.e.}, we do not observe UCEL in O-up.
Interestingly, for O-down, light emission from the neutral molecule ($Q$, $\hbar\omega \approx$ 1.80\,eV) becomes visible already at $V_\mathrm{b}$ $\leq$ -1.35\,V, which coincides with the onset of the PIR (see Fig.~\ref{fig:Electronic_struc}b and d). This UCEL is comparable in intensity to the $Q$ band at electron energies above the photon energy and is intrinsically only possible if two electrons contribute to the emission of a single photon. Light emission from the positively charged molecule in O-down configuration ($X^+$, $\hbar\omega \approx$ 1.45\,eV) only appears for $V_\mathrm{b}$ $\leq$ -1.85\,V, similarly to O-up (see Fig.~\ref{fig:Electronic_struc}c). This means that flipping VOPc is gating UCEL, although the level alignment (energy of PIR) and the spacing of the emitting levels (both $Q$ and $X^+$) are almost identical for the two orientations. 
The intensities of the $X^+$ and $Q$ bands increase further for $V_\mathrm{b}$ < -2.2\,V for both O-up and O-down orientations, indicating the onset of additional excitation pathways \cite{chen_spin-triplet-mediated_2019}.

To further corroborate the presence of two-electron processes leading up to the UCEL in the O-down, we also measured the dependence of the photon count rate on the tunneling current \cite{doppagne_electrofluorochromism_2018, rai_boosting_2020,chen_spin-triplet-mediated_2019}.
A double logarithmic plot (see Fig.~\ref{fig:Electronic_struc}e and f) shows that the photon count rate is proportional to $I^\alpha$ with 1 < $\alpha$ < 2 for both configurations, but closer to two for O-down. This is further confirmation of a multi-electron process via an intermediate, long-lived state in O-down. In full agreement, a time-resolved measurement, second-order correlation function (g$^{(2)}(\tau)$), shows that the overall excitation-emission cycle is slower for O-down compared to O-up (see SI Note 3).

\begin{figure}[!h]
\centering
\includegraphics[width = 1\columnwidth]{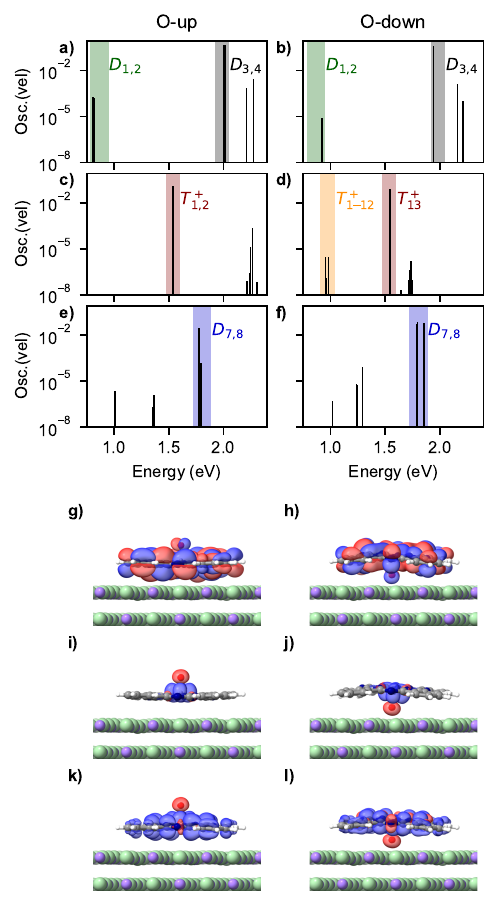}
\caption{\label{fig:ExcitationEnergies} {\bf Calculated TD-PBE0  excitation spectra and NTOs} (a) and (b) Oscillator strength in the velocity representation of the lowest excitations of the ground state in the duplet spin configuration of neutral molecules in O-up, O-down, respectively. (c) and (d) Oscillator strength of the lowest excitations of the positively charged molecule in the triplet spin configuration of O-up, O-down, respectively. Note that for the O-down configuration, there are additional low-energy states (yellow band). (e) and (f) Oscillator strength of the lowest second excitations starting from $D_{1,2}$ in the Duplet spin configuration of O-up, O-down, respectively, as obtained from quadratic response TD-DFT. Colored excitation bands in (a)-(f) that are referred in the text are indicated as a guide to the eye and labeled according to the spin configuration of the excited state. (g) and (h) Side view of particle NTO of the $D_{1,2}\rightarrow D_0$ transition of O-up, O-down orientations. (i) and (j) Calculated spin density of O-up and O-down in the neutral molecule. (k) and (l) Calculated spin density of O-up and O-down in the cation.
} 

\end{figure}

To understand the microscopic mechanism that leads to the gating of the UCEL by the adsorption configuration of VOPc, we performed extensive DFT calculations of VOPc adsorbed on NaCl. 
The calculated TD-PBE0 \cite{Perdew.Burke.ea:Generalized.1996,Adamo.Barone:Toward.1999} and TD-CHYF \cite{Holzer.Franzke:General.2025} excitation spectra allow us to identify the experimentally observed emission bands $Q$ of both orientations with $D_{3,4} \rightarrow D_0$ transitions of VOPc in a doublet spinstate (grey shaded bands in Fig.~\ref{fig:ExcitationEnergies}a and b) \cite{kaiser_single-molecule_2019,debnath_triggering_2022}. $D_{3,4}$ hereby refers to the third and fourth excited state, which are nearly degenerate. The used labeling is analogous to standard TD-DFT based on triplet ground states.
Note that there is a consistent overestimation of excitation energies in DFT by $\approx$ 0.2 eV. The reorganization of NaCl only has a slight effect on the calculated excited state energy levels, which we have calculated to be less than 0.02 eV, and therefore insignificant when compared to the overall TD-DFT error. However, we expect the relaxation of VOPc on the surface to be considerably more important for the electronic structure than the slight alterations of the NaCl surface in the vicinity of the VOPc unit. Generally, the changes in the electronic structure induced by deformations of the NaCl layers are expected to be well below the intrinsic accuracy of TD-CHYF, which is found to be accurate to 0.1-0.2\,eV. For both orientations, the lowest-lying excited states $D_{1,2}$ at 0.8 to 0.9 eV (TD-PBE0) and 1.0 eV (TD-CHYF) are unlikely to decay via photon emission, as indicated by very low oscillator strength (green-shaded bands in Fig.~\ref{fig:ExcitationEnergies}a and b), and are outside of the energy window of our detector (see SI Fig.~S4). The experimentally observed $X^+$ line can be explained by transitions of the positively charged VOPc in a triplet spin configuration with two unpaired electrons (brown-shaded band in Fig.~\ref{fig:ExcitationEnergies}c and d) at $\approx$ 1.5 eV, in full agreement with previous work \cite{debnath_triggering_2022}. Surprisingly, O-down shows additional 'dark' states with very low oscillator strength $T_{1-12}$ (yellow-shaded band in Fig.~\ref{fig:ExcitationEnergies}d) that are fully absent in O-up.
The excitation energies of the singlet spin configuration of the cation do not match with our experiment (see SI Fig.~S4). In addition to the excitations of the ground state, we also managed to calculate the excitation spectrum of the $D_{1,2}$ states using quadratic response TD-DFT via the PBE0 functional \cite{Himmelsbach.Holzer:Excited.2024} (see Fig.~\ref{fig:ExcitationEnergies}e and f). The summed oscillator strengths for the lowest excited states excitations within the energy range of the experiment compare as 0.059 to 0.204
for the O-up and O-down configurations, respectively, and are dominated by the $D_{1,2} \rightarrow D_{7,8}$ transitions (blue shaded band in Fig.~\ref{fig:ExcitationEnergies}e and f).
The excitations in VOPc are of multi-electron nature and can best be depicted by their natural transition orbitals (NTO). Virtual NTOs of the excited states $D_{1,2}$ do not differ much for the two orientations except for a slightly increased overall distance from the substrate (see Fig.~\ref{fig:ExcitationEnergies}g and h). However, the location of the unpaired electron, described by the calculated spin density, is significantly different for O-up and O-down \cite{malavolti_tunable_2018}, as is shown for the neutral molecule and the cation in Figs. \ref{fig:ExcitationEnergies}i, j and \ref{fig:ExcitationEnergies}k, l respectively.

\begin{figure}[h!]
\centering
\includegraphics[width = 1\columnwidth]{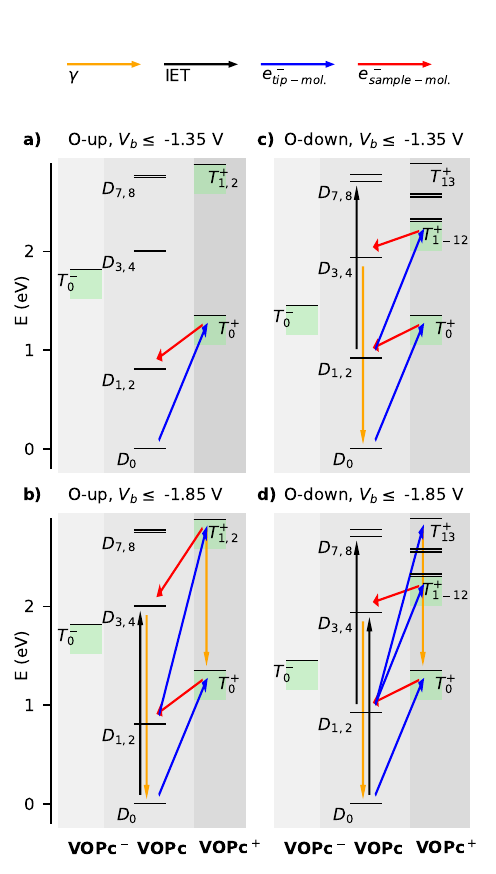}
\caption{\label{fig:Model} {\bf Many-body picture of excitation-emission cycles in VOPc} Transitions under photon emission are marked by yellow arrows, inelastic energy transfer (IET) is marked by black arrows, charging via electron transfer from the molecule to the tip and from the sample to the molecule is marked by blue and red arrows, respectively. Important processes in O-up at (a) -1.85 < $V_\mathrm{b}$ < - 1.35\,V and (b) $V_\mathrm{b}$ < 1.85\,V. Important processes in O-down at (c) -1.85 < $V_\mathrm{b}$ < - 1.35\,V and (d) $V_\mathrm{b}$ < 1.85\,V. 
Level spacing drawn according to DFT calculation, energy shift between charged states extracted from experiment (VOPc$^+$ from onset of PIR at $\approx 1.35\,V$, VOPc$^-$ from onset of NIR at $\approx 1.6\,V$). In c and d, 12 triplet states T$_1^+$-T$_{12}^+$ are found for VOPc$^+$ for 
the O-down configuration, where in-between states are omitted for clarity.}
\end{figure}

These DFT calculations allow us to explain the experimentally observed gating of the UCEL by the adsorption configuration and to describe the excitation-emission cycle for the two orientations in a comprehensive many-body picture (see Fig.~\ref{fig:Model}) \cite{jiang_many-body_2023,hung_bipolar_2023,friedrich_fluorescence_2024,kaiser_electrically_2025}.

For both O-up and O-down orientations, $D_{0} \rightarrow T_{0}^+$ sets in at $V_\mathrm{b}$ = -1.35\,V via removal of an electron from the molecule to the tip, as can be inferred from the experimentally observed onset of the PIR (see Fig.~\ref{fig:Electronic_struc}a and b and lower blue arrows in Fig.~\ref{fig:Model}a and c). Subsequent electron capture from the sample can bring the molecule to $D_{1,2}$ (see red arrows in Fig.~\ref{fig:Model}a and c) in both orientations. However, only in the O-down orientation, state reordering leads to low-energy triplet states, which allows $D_{1,2} \rightarrow T_{1-12}^+$ via CI (see upper blue arrow in Fig.~\ref{fig:Model}c). This step is impossible in the O-up orientation because the lowest triplet states are at too high energies. The UCEL process in O-down is completed by subsequent electron capture $T_{1-12}^+ \rightarrow D_{3,4}$ and recombination of $D_{3,4}\rightarrow D_{0}$ under photon emission (see upper red and yellow arrows in Fig.~\ref{fig:Model}c). This explains the observed onset of light emission (see Fig.~\ref{fig:Electronic_struc}d) by purely CI-mediated UCEL via $D_{0} \rightarrow T_{0}^+ \rightarrow D_{1,2} \rightarrow T_{1-12}^+ \rightarrow D_{3,4}$ (see Fig.~\ref{fig:Model}c), which involves only the positively charged state of the molecule. The direct assignment of the PIR/NIR with the energy of the triplet states is an overestimation of these energies because it neglects the voltage drop in the NaCl \cite{kaiser_charge-state_2023} and the reorganization energy \cite{fatayer_reorganization_2018, scheuerer_charge-induced_2019}. Assuming a reorganization energy of $\approx$ 0.8eV, a systematic study on ZnPc variants revealed differences between the apparent dI/dV gap and the electronic gap of 0.2 to 0.24 eV \cite{vasilev_exploring_2024} and recent experiments on free base porphyrin and Pc molecules found the PIR to be about 0.3 eV higher than the energy of the corresponding charge state \cite{ammerman_tunable_2026}. We therefore approximate this overestimation by 0.3 eV as indicated by light green shaded boxes in our model (see Fig.~\ref{fig:Model}). This energy correction does not affect our explanation of possible transitions as described above. Note that the NIR at an onset of 1.6 eV is inaccessible from $D_{1,2}$, because electron capture from the substrate provides only minimal energy (corresponding to the small voltage drop across NaCl \cite{kaiser_charge-state_2023}). The transition from $D_{3,4}$ to the NIR is a possible quenching path of Q emission, but does not influence the overall picture. Also note that light emission from $T_{1-12}^+ \rightarrow D_{0}$ is not expected from low oscillator strengths and lies outside the range of our detection system.
Compared to direct excitation via CI, indirect effects such as inelastic energy transfer (IET) are typically less likely \cite{luo_anomalously_2024}. Although there is no experimental indication for IET, in principle, it would allow the molecule to be pumped from $D_{1,2}$ to higher excited states. 
Comparing the oscillator strengths for transitions between excited-states of O-up and O-down in our DFT calculations shows that $D_{1,2} \rightarrow D_{7,8}$ are three times less likely in O-up than in O-down (see Fig.~\ref{fig:ExcitationEnergies}e and f) and transitions to any other higher lying states are very unlikely in both configurations. More generally, the reduced distance of the Pc macrocycle to the substrate in O-up (see Fig.~\ref{fig:ExcitationEnergies}g and h), the shorter life-time of $D_{3,4}$ and the shorter overall excitation-emission cycle (see second-order photon correlation measurement (g$^{(2)}(\tau)$) in SI Fig.~S3) might point toward shorter lifetime of $D_{1,2}$ in O-up compared to O-down, which would render a second tunneling process less likely. However, several other factors such as deformation of the molecular orbitals or changes in the overlap integrals among frontier molecular orbitals and substrate wave functions—could also might influence the coupling. Note that the charge state lifetime is of the order of a few hundred ps and might well contribute equally to the overall excitation-emission cycle.
This fully explains why UCEL is quenched in O-up, that is, it does not show any light emission at -1.85\,V < $V_\mathrm{b}$< -1.35\,, neither the Q band nor the $X^+$ band.
The absence of triplet emission at voltages -1.85\,V < $V_\mathrm{b}$ < -1.45\,V for both orientations indicates that the $T_0^+$ lifetimes are too short to allow for a second excitation step via IET. This is in agreement with the low efficiency of IET in general and short lifetimes of charged states of Pc molecules on 3 ML NaCl (tens to hundreds of ps) \cite{kaiser_charge-state_2023} compared to the average time elapsed between two consecutive tunneling electrons at the low currents used in our experiment (typically a few ns).

At $V_\mathrm{b}$ = -1.85\,V, $D_{0}\rightarrow D_{3,4}$ becomes energetically possible via IET (see black arrows in Fig.~\ref{fig:Model}b and d). This excitation via IET does not involve any transiently charged states and is thus much faster than CI, which would explain the absence of an anti-bunching dip on the time-scale of 0.5 ns (see SI note 3). The simultaneous onset of $Q$ ($D_{3,4} \rightarrow D_{0}$, left yellow arrow in Fig.~\ref{fig:Model}b) and more intense $X^+$ ($T_{1,2}^+ \rightarrow T_{0}^+$, right yellow arrow in Fig.~\ref{fig:Model}b) emission bands in O-up (see bias dependence in SI Note 2) suggest a CI-mediated process: $T_{1,2}^+$ is reached via electron removal from $D_{1,2}$ (upper blue arrow in Fig.~\ref{fig:Model}b) and $D_{3,4}$ is reached via electron recapture from the substrate (upper red arrow in Fig.~\ref{fig:Model}b). According to our DFT calculations and the experimentally observed onset of the PIR, $D_{1,2} \rightarrow T_{1,2}^+$ via CI becomes energetically possible starting from $V_\mathrm{b}$ $\approx$ 2.0\,eV ( = 1.35 eV + 1.45 eV - 0.8 eV) which agrees with the observed threshold voltage of 1.85 V within the error of DFT and our experimental values. The observed current dependency shown in Fig.~\ref{fig:Electronic_struc} suggests that both single-electron processes, like IET, and two-electron processes contribute to the excitation. Both processes, IET and CI via the short-lived $T_{1,2}^+$ (see high oscillator strength), are less likely than the CI mechanism via non-radiating states in O-down, which explains the lower intensity of Q emission in O-up.
An analogous mechanism explains the onset of $X^+$ at $V_\mathrm{b}$= -1.85\,V in O-down by $D_{1,2} \rightarrow T_{13}^+ \rightarrow T_{0}^+$. 
At similar energies, the singlet cation becomes accessible and might also allow for the Q emission (see SI note 5).
At higher voltages, additional excitation pathways via CI may open up in both configurations, further increasing the radiative decays.
Note that we restrict this model to the electron configurations that are necessary to explain the experimental data and refrain from discussing the possibility of additional spin states \textit{e.g.} singlet charged state, neutral state with three unpaired spins.

\section{Conclusion}
We have experimentally and theoretically shown that UCEL can be gated by the adsorption configuration of VOPc on NaCl, with UCEL possible in the O-down configuration, but not in the O-up configuration. In the O-down configuration, VOPc shows UCEL via charge injection, similar to recent work on UCEL in free base Pc \cite{luo_anomalously_2024}. While the model explaining the latter relies on both anionic and cationic states, the analogous mechanism is energetically not accessible for VOPc/NaCl/Au(111). Instead, state reordering in VOPc O-down lowers the excited triplet states of the cation, thereby enabling UCEL via two cationic states. These states correspond to excitation from the HOMO-N to the single occupied orbital. In the O-up configuration, the excited states of the cation are too high in energy to allow for UCEL. The origin of this state reordering critical for the cationic UCEL can be rationalized by the location of the unpaired spin at the Vanadyl. Depending on the orientation of the molecule, this leads to very different distances of the oxygen to the underlying substrate, and thus to different screening. This manifests itself in the experiment as the absence of UCEL in O-up and the absence of lower-lying triplet states in the DFT calculation. 
Typically, coupling to the substrate in most STML studies mainly shifts the ion resonances but barely affects the excitation energies of the adsorbed molecule. Similarly, the basic properties of Pc radicals with an unpaired spin seem to be mostly unchanged by the interaction with the substrate \cite{malavolti_tunable_2018, atzori_room-temperature_2016,vasilev_internal_2022,reecht_-radical_2019}, with emission lines shifting by a few tens of meV \cite{dolezal_mechano-optical_2020}. Our comparison of O-up and O-down VOPc, however, indicates that the interaction of an unpaired spin with the underlying substrate can indeed have a profound influence on the molecular state reordering, opening new pathways for tuning molecular electroluminescence.

\section{Acknowledgment}
C.H. and C.R. acknowledge funding by the Volkswagen Foundation. W.W. acknowledges funding by the Deutsche Forschungsgemeinschaft (DFG, German Research Foundation) through the Collaborative Research Center “4f for Future” (CRC 1573, project number 471424360) project C1 and support by the Helmholtz Association via the programs Natural, Artificial, and Cognitive Information Processing (NACIP) and Materials Systems Engineering (MSE).

\clearpage

\renewcommand{\figurename}{Fig. S}

\setcounter{figure}{0}
\onecolumngrid
\setcounter{section}{0}

\section*{Supplementary Information}
\maketitle
\section{Note 1: Experimental setup and sample preparation}

Low-temperature STM measurements were carried out using a custom-built STM operating at $\approx$ 4.4\,K under ultra-high vacuum conditions ($\approx$ $10^{-10}$ mbar). The setup provides optical access, enabling efficient light collection from the STM junction \cite{edelmann_light_2018s}. Electroluminescence spectra were recorded using a spectrometer (150 mm focal length) equipped with a grating of 300 groves mm$^{-1}$. The optical resolution of this setup is about 2\,nm when the entrance slit is closed to 10\,$\mu$m. All photon spectra presented in the manuscript are corrected for the collection efficiency of the detection setup except for the losses in the parabolic mirror which are unknown. To measure the second-order correlation function (g$^{(2)}(\tau)$), a Hanbury-Brown-Twiss interferometer was used with two single photon avalanche diodes (SPAD) (Excelitas SPCM-AQRH-14TR-BR1) connected to a time tagger (qutools: quTAG Standard).

An Au(111) and a Ag(111) single crystals (from MaTeck) were cleaned by repeated cycles of Ar+ sputtering and subsequent annealing at temperatures between 750-850\,K. First, NaCl was thermally evaporated onto the cleaned Au(111) surface while keeping the substrate at $\approx$ 450\,K to promote the formation of bi-and trilayers. A thick layer of VOPc molecules (several monolayers) was then deposited by thermal sublimation from a Kentax 3-cell evaporator at 550\,K on a clean Ag(111) substrate. Subsequently, this sample was transferred to the STM chamber and placed in front of the actual NaCl/Au(111) sample sitting at the cooling stage. By flash annealing the Ag(111) with the thick film of VOPc, it served as the source of molecules that were deposited onto the NaCl/Au(111). This allowed us to perform the deposition with minimal substrate temperature and a minimal amount of molecules sublimed in the STM chamber.

\begin{figure}[ht]
\centering
\includegraphics{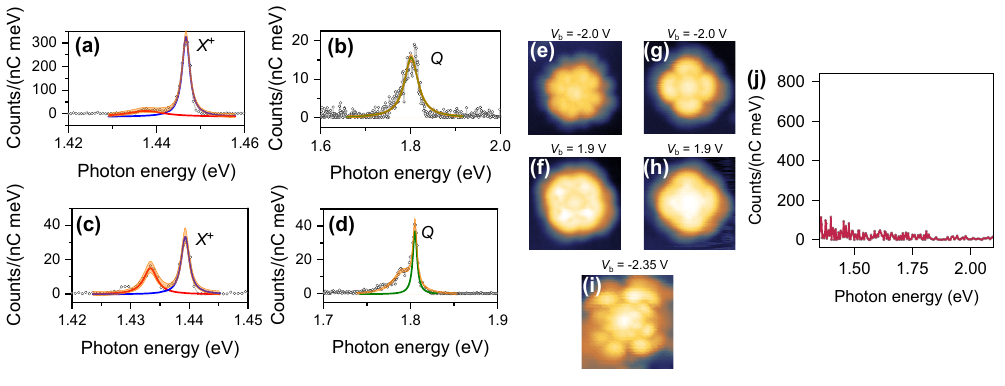}
\caption{\label{fig:VPc} \textbf{STML spectra recorded on VOPc in} (a)-(b) O-up and (c)-(d) O-down configurations. Experimental data are shown in black, whereas the colored solid curves show the Lorentzian fit of the charged (blue and red) and the neutral (dark green) emission lines. The orange shaded region in all the spectra represents the 95\% confidence interval of the fit. (e) and (f) show the topographies recorded on a single VOPc (O-up) at $V_\mathrm{b}$ = -2.0\,V, $I$ = 1\,pA and $V_\mathrm{b}$ = 1.9\,V, $I$ = 1\,pA. (g) and (h) show the topographies of VPc at the same $V_\mathrm{b}$ and current as for e and f, respectively. (i) Topography of VOPc in O-down configuration ($V_\mathrm{b}$ = -2.35\,V, $I$ = 4\,pA). (j) STML spectrum recorded at VPc ($V_\mathrm{b}$ = -2.7\,V, and $I$ = 30.5\,pA, $t_{exp}$ = 5\,s). Topographies (e-h) are 4.5 \,nm$\times$ 4.5\,nm and 2.5 \,nm$\times$ 2.5\,nm in i.}
\end{figure}

\section{Note 2: Additional STML spectra from VOPc and VPc}
High resolution spectra recorded with the spectrometer entrance slit opening of 10\,$\mu$m are shown in Fig. S\ref{fig:VPc}a. Side peaks of X$^+$ and Q lines are clearly visible for both configurations. The peak widths of both emission lines, charged ($X^+$) and neutral ($Q$) in the O-up (Fig. S\ref{fig:VPc}(a), (b) and O-down (Fig. S\ref{fig:VPc}(c), (d) configurations are determined by fitting with a Lorentzian line shape. The full width half maxima of the $Q$/$X^+$ line is $\approx$\,40\,meV/2\,meV and $\approx$\,5.8\,meV/2\,meV in O-up and O-down configurations, respectively.

The removal of the oxygen atom from the VOPc molecule was also possible. This was typically achieved by scanning the molecule at higher bias voltages ($V_\mathrm{b}$ > 2.5\,V). This change was evident both in the STM topography and in the STML properties. Fig. S\ref{fig:VPc}e-i show the topographies of VOPc (O-up and O-down) and VPc, showing clearly distinguishable structural differences. Furthermore, the STML spectrum recorded over the macrocycle of the VPc molecule exhibits no distinct peaks associated with any electronic transitions (see Fig. S\ref{fig:VPc}j), in complete agreement with previous work \cite{kaiser_single-molecule_2019s}.  

\begin{figure}[ht]
\centering
\includegraphics{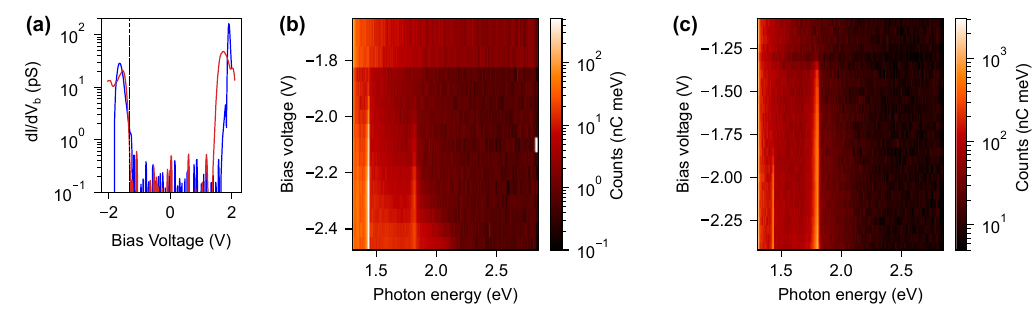}
\caption{\label{fig:Bias_dep_color} (a) d$I$/d$V_\mathrm{b}$ for O-up (blue) and O-down (red) plotted in log scale to highlight the onset of the PIR. The vertical dashed line is at -1.35\,V. (b) STML spectra recorded as a function of applied bias voltages for VOPc in O-up configuration ($I$ = 40\,pA, $t$ = 30\,s) (c) VOPc in O-down configuration ($I$ = 45\,pA, $t$ = 10\,s).}
\end{figure}

Furthermore, to clearly show the emergence of the UCEL in case of VOPc O-down configuration compared to O-up, color map of bias dependent measurement over wide bias voltage range is shown in Fig. S\ref{fig:Bias_dep_color}b and c. 

\begin{figure}[ht]
\centering
\includegraphics{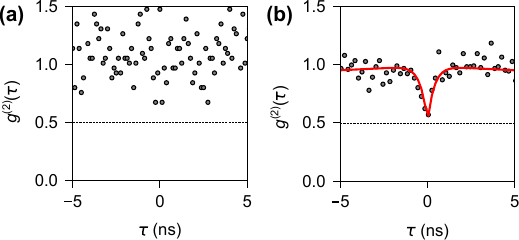}
\caption{\label{fig:G2} Second order photon correlation measurement (g$^{(2)}(\tau)$) recorded on VOPc in (a) O-up configuration and (b) O-down configuration ($V_\mathrm{b}$ = -2.35\,V, and $I$ = 450\,pA). Integration time for both are 200\,s.}
\end{figure}

\section{Note 3: Photon correlations for O-up and O-down VOPc }
For the time correlation measurements of the emitted photons from the VOPc molecules, a home built Hanbury Brown–Twiss (HBT) interferometer was used. Two single-photon avalanche diodes (SPADs) are used for the photon statistics with a jitter of 250\,ps. The histogram was recorded with 0.2 \,ns bin width over the period of 200\,s. The second-order correlation function (g$^{(2)}(\tau)$) measured on the two VOPc orientations are shown in Fig. S\ref{fig:G2}a (O-up) and Fig. S\ref{fig:G2}b (O-down). In the histogram of the O-up configuration, no dip at $\tau$ = 0\,s was observed, however, for the O-down configuration, a clear dip corresponding to non-classical photon behavior, antibunching is observed. The histogram shown in Fig. S\ref{fig:G2}b was normalized and fitted with two independent exponential decays \cite{khasminskaya_fully_2016s}:

\begin{equation}
g^{(2)}(\tau) = 1 - a e^{-\gamma_1 |\tau - \tau_0|} - b e^{-\gamma_2 |\tau - \tau_0|}
\end{equation}

Here, the constants $a$ (negative) and $b$ (positive) correspond to (contributions from bunching and antibunching ($\tau_0$ is center of the antibunching). The time constant $\gamma_2^{-1}$ obtained from the fit is 0.38$\pm$0.11\,ns, which corresponds to the charge-state lifetime of the system that is, the time for the entire excitation-emission cycle and not just the excited-state lifetime \cite{kaiser_electrically_2025s}. Within our measurement time window, O-down shows a longer excitation-emission cycle compared to the O-up configuration. The time constant obtained from the VOPc O-down is in a similar range as for the other phthalocyanine molecules in similar junctions \cite{zhang_electrically_2017s,kaiser_electrically_2025s}.

\begin{figure}[h!]
\centering
\includegraphics[width = 0.5\columnwidth]{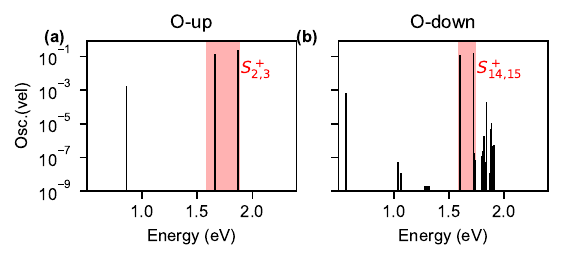}
\caption{\label{fig:AddExcitationEnergies} Excitation energies of VOPc$^+$ in singlet spin configuration. The theoretically expected emission band at around 1.7 eV is not observed experimentally. Note that the transitions for O-down at around 1.3 eV show oscillator strengths of less than $10^{-8}$ in the calculations and are given a finite value for visualization only.}

\end{figure}

\section{Note 4: Density functional theory calculations}
A Gaussian-type local basis (def2-TZVP) type was used for the elements C, H, N, and V, while def2-SVP was used for Na and Cl.
\cite{Weigend.Ahlrichs:Balanced.2005s}
For the calculation on sodium chloride (NaCl), three layers were modeled by using crystal parameters for NaCl, while the VOPc
structure was fully relaxed using the PBE functional
\cite{Perdew.Burke.ea:Generalized.1996s}.
Reorganization effects including the change of the NaCl
surface are found to play only a minor role on the overall 
observed behavior of the molecular systems.
The excited states themselves do barely change upon
a fully relaxed geometry, deviating only by 0.01-0.02~eV 
from the respective values obtained with the NaCl surface
being assumed to be rigid. Our model therefore remains virtually unchanged, and these relaxation
effects are well below the intrinsic accuracy of TD-CHYF, 
which is found to be accurate to 0.1-0.2~eV \cite{Holzer.Franzke:General.2025s}.
For the DFT part, the ground state energies were converged to 10$^{-8}$ Hartree, and the residual norm to 10$^{-7}$. All DFT calculations were performed using the resolution-of-the-identity (RI) approximation for the Coulomb part, and the semiJK algorithm for the range-separated exchange part.\cite{holzer_improved_2020s} For integrating the DFT exchange-correlation part a DFT grid of size 3 was used, while for the range-separated exchange a "fine" grid was used. All calculations were performed using a development version of the \textsc{Turbomole} package 
 \cite{balasubramani_turbomole_2020s,Franzke.Holzer.ea:TURBOMOLE.2023s}.
 Excited state spectra were obtained from the optimized structure by performing time-dependent density functional theory (TD-DFT) calculations using the hybrid functional PBE0.
 \cite{ Adamo.Barone:Toward.1999s} Accordingly, state-to-state
 transition probabilities were calculated using quadratic response
 PBE0, and yielded notable transition probabilities between the
 $D_{1/2}$ and $D_{7/8}$ excited states, while other transitions from
 $D_{1/2}$ are predicted to be dark. Linear response TD-DFT 
 results were checked by the CHYF local hybrid functional,\cite{Holzer.Franzke:General.2025s} 
 which is capable of describing long-ranged charge-transfer excitations. 
 Due to no available implementation for quadratic response 
 TD-DFT of local hybrid functionals like CHYF, PBE0 results are reported throughout 
 the manuscript.

\section{Note 5: Excitation by charge injection via the Singlet cation}
\begin{figure}[h!]
\centering
\includegraphics[width = 0.5\columnwidth]{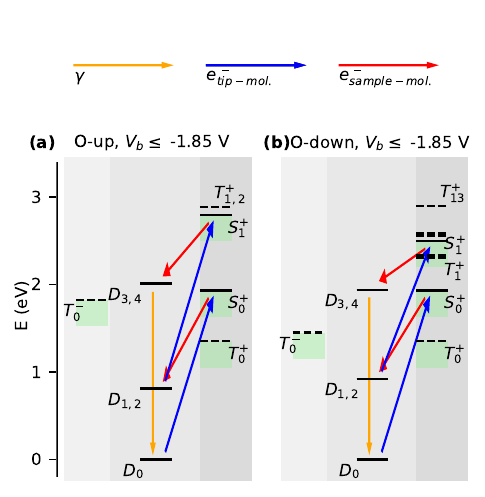}
\caption{ Excitation via charge injection considering the pathway via the singlet cation for a) O-up and b) O-down. Triplet states are indicated by dashed lines for comparison. The overestimation of the charged states by about 0.3 eV is indicated by light green boxes, but omitted for the higher triplet states for simplicity.} 
\label{fig:Charge injection via singlet}
\end{figure}

In the positively charged state, triplet and singlet spin configurations are possible. Our CHYF calculations outline that the triplet state is more stable than the singlet state, with the stabilization energy being 49~kJ/mol (=0.49~eV) for the O-up and 56~kJ/mol (=0.58~eV) for the O-Down configuration. The electron energy required for charge injection via the singlet cation can be estimated from the PIR and this stabilization energy to 1.35 eV + 0.58 eV = 1.93 eV, which already exceeds the experimentally determined energy of the Q emission band. Note that the PIR is an overestimate of the energy of the singlet cation, which we approximate by 0.3 eV based on recent studies \cite{vasilev_exploring_2024s,ammerman_tunable_2026s}.
The singlet configuration potentially serves as third relay state at lower energies: $D_{0} \rightarrow T_{0}^+ \rightarrow D_{1,2}\rightarrow S_{1}^+\rightarrow D_{3,4}$ (see Fig. S\ref{fig:Charge injection via singlet}). However, for O-up, the common onset of emission from the neutral doublet and the cation in triplet configuration suggests that the singlet is only involved at higher energies. For O-down, the first excited state of the singlet lies closely spaced above the lowest excited triplet states and thus might well contribute to UCEL at higher energies. At energies higher than 1.93 eV, excitation of $D_{3,4}$ becomes possible without involving triplet states via $D_{0} \rightarrow S_{0}^+ \rightarrow D_{1,2}\rightarrow S_{1}^+\rightarrow D_{3,4}$ (see Fig. S\ref{fig:Charge injection via singlet}). Clearly, this can not explain the experimentally observed UCEL with an onset at -1.35 eV. Note that in our model depicted in Fig. S\ref{fig:Charge injection via singlet}, energies are taken from DFT except for the onset of the PIR at 1.35 eV. Also note that, the energy of the PIR equals the energy of the respective spin state plus the energy required for charging, including the reorganization energy.

\section{Note 6: Comparison with previous work on UCEL in a non-radical molecule}
\begin{figure}[h!]
\centering
\includegraphics[width =0.5\columnwidth]{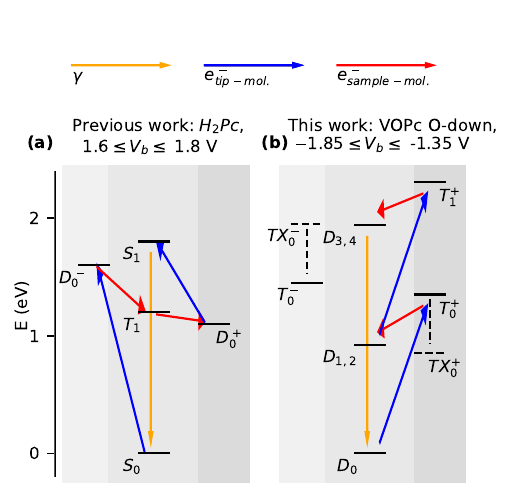}
\caption{\label{fig:SI_Model} {\bf Many-body picture of UCEL in H$_2$PC/3ML NaCl/Au(111) and  VOPc/3ML NaCl/Au(111) analogously to Fig.~4 and based on previous work \cite{jiang_many-body_2023s}} Transitions under photon emission are marked by yellow arrows, charge injection via electron transfer between the molecule and the tip and between the sample and the molecule is marked by blue and red arrows, respectively. (a) UCEL in previous work \cite{luo_anomalously_2024s} for 1.6 < $V_\mathrm{b}$ < 1.8\,V based on the following values from \cite{luo_anomalously_2024s}: $D_0^-:\,1.6\, eV$, $T_1:\,1.2\, eV$,$D_0^+:\,1.1\, eV$,$S_1:\,1.8\, eV$  (b) UCEL in this work -1.85 $\leq V_\mathrm{b} \leq$- 1.35\,V.}
\end{figure}

Electrical up-conversion in systems similar to the one described here requires net transport of (at least) two charges from the tip to the sample via the molecule. This can be divided into four sequential transitions from the ground state via three relay states, alternating between neutral and transiently charged species \cite{kaiser_charge-state_2023s}. \\
The UCEL model previously proposed by Luo et. al. \cite{luo_anomalously_2024s} is transferred to the many-body description as shown in SI Fig. S\ref{fig:SI_Model}a. Luo et. al. report efficient up-conversion in H$_2$PC/3ML NaCl/Au(111) via $S_{0} \rightarrow D_{0}^- \rightarrow T_{1}\rightarrow D_{0}^+\rightarrow S_{1}$ (Fig.~2b in \cite{luo_anomalously_2024s} and description therein). For a side-to-side comparison, Fig. S\ref{fig:SI_Model}b shows a copy of Fig.4 in our main text based on the suggested excitation via $D_{0} \rightarrow T_{0}^+ \rightarrow D_{1,2} \rightarrow T_{1}^+ \rightarrow D_{3,4}$.
In addition to the apparent fact that states of different spin multiplicities are involved, the model in Fig. S\ref{fig:SI_Model}a) relies on both cationic and anionic states, while Fig. S\ref{fig:SI_Model}b) only involves cationic states. This entails two main differences:\\

First, an excitation process analogously to (a) would be impossible to achieve via engineering of level-alignment in systems where the energies of the ion resonances are too large compared to energies of the neutral states, such as in the present case of VOPc: The analogous mechanism would be UCEL via $D_{0} \rightarrow T_{0}^- \rightarrow D_{1,2} \rightarrow T_{0}^+\rightarrow D_{3,4}$. This would require $D_{1,2} <  T_{0}^- < D_{3,4}$ and $D_0 < T_{0}^+ < D_{1,2}$ (alignment of levels analogous to (a)). While the first condition is met for VOPc/NaCl/Ag(111), the second condition would require a suitable substrate/workfunction that shifts $T_{0}^+$ by about 0.5 eV to lower energies such that $D_{1,2} \rightarrow T_{0}^+$ becomes possible via charge injection from the substrate (see dashed lines and $TX_0^+$ in Fig. S\ref{fig:SI_Model}b)). However, at the same time, $T_0^-$ would be shifted to higher energies above $D_{3,4}$ (see dashed lines and $TX_0^-$ in Fig. S\ref{fig:SI_Model}b)), and UCEL would be impossible. The mechanism proposed in this paper is possible because it relies on two cationic states, such that the alignment of the NIR becomes irrelevant. \\
Under typical tunneling conditions, and in particular for low set point currents, the vacuum barrier to the tip is much larger than the barrier formed by 3 ML NaCl \cite{kaiser_charge-state_2023s} and the corresponding rates for charge injection from the tip (blue arrows in our models in Fig.~4, Fig. S\ref{fig:SI_Model} and Fig. S\ref{fig:Charge injection via singlet}) are typically smaller than the rates for charge injection from the substrate (red arrows in our models in Fig.~4 and Fig. S\ref{fig:SI_Model}, Fig. S\ref{fig:Charge injection via singlet}).
Considering this difference, we can compare requirements for the lifetimes of the relay states:
In both models, the first relay state ($D_{0}^- $ in (a), $T_{0}^+$ in (b) has to be long-lived compared to charge injection from the substrate (red arrows). However, the requirements for the lifetimes of the second and third relay states in a) and b) are swapped: In a), the second relay state, $T_{1}$, has to be long-lived enough to allow a second charge injection from the substrate, while in b) $D_{1,2}$ has to be long-lived enough to allow charge injection from the tip. In a), the third relay state, $D_{0}^+$, has to be long-lived enough to allow a second charge injection from the tip, while in b) $T_{1}^+$ has to be long-lived enough to allow charge injection from the substrate.

\section{Note 7: Projection of many-body states onto single particle wavefunctions}

\begin{figure}[h!]
\centering
\includegraphics[width = 0.95\columnwidth]{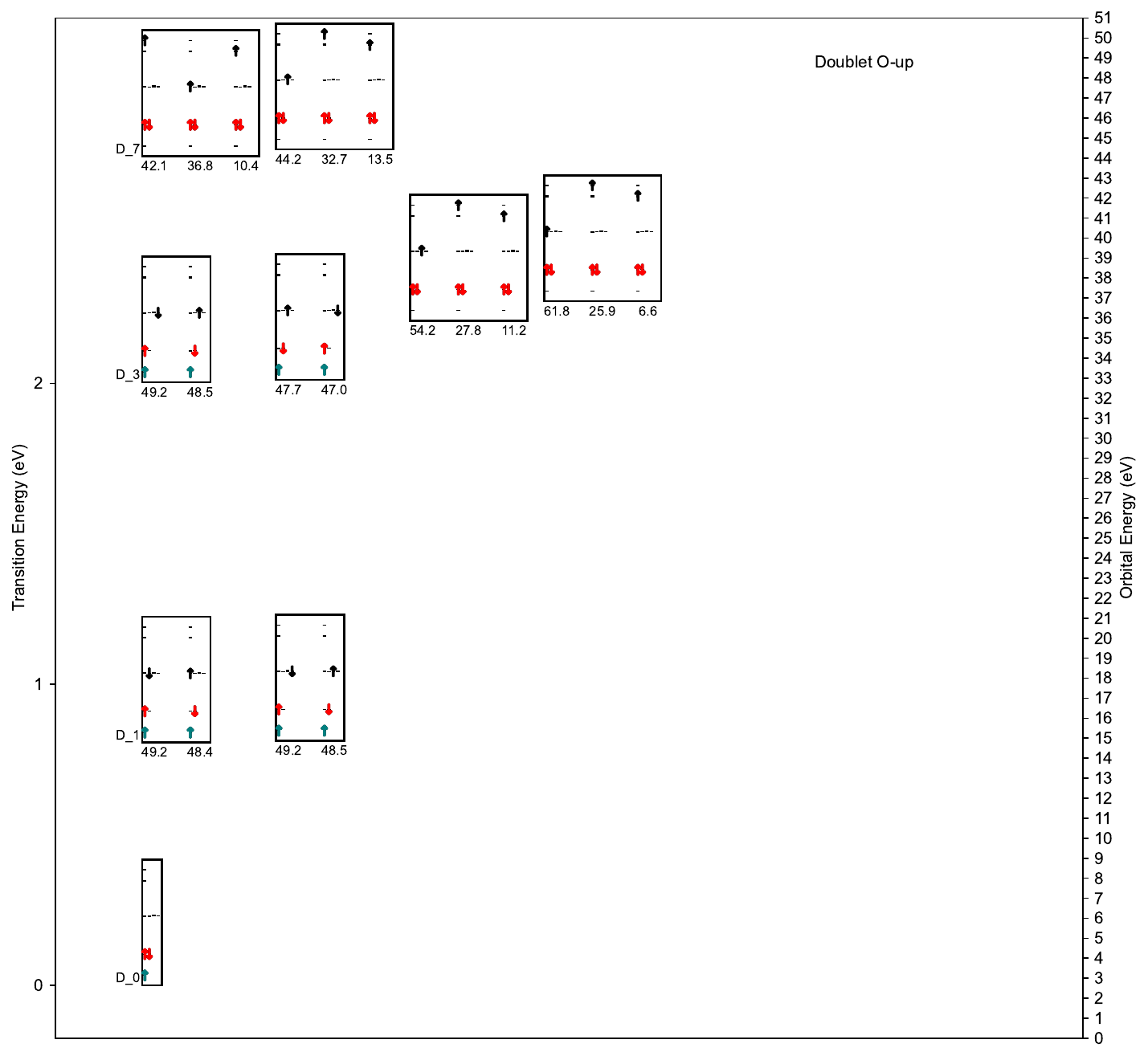}
        \caption{\textbf{Many-body states represented by single particle wavefunctions:} Many-body states for the neutral doublet spin configurations for O-up . For each many-body state, a box comprises the dominant single particle wavefunctions contributing to the different excitations (relative contributions are indicated below the box). Energies of single particle wavefunctions (small horizontal dashes) are plotted on a consistent energy scale (y axis on the right). Energies of many-body states are scaled by a factor of 18 (y axis on the left). Single particle avefunctions 3171beta and 3172alpha are depicted in red, 3171alpha is depicted in teal. }
\label{fig:SI-Duplett_normal_ind}
\end{figure}

Figures S~\ref{fig:SI-Duplett_normal_ind} - S\ref{fig:SI-Singlett_reverse_ind} show a projection, where we have assigned changes in the excitation vector to the corresponding
one-particle state originating from the Kohn--Sham eigenstates. The eigenstates obtained from the time-dependent DFT problem
\cite{Stratmann.Scuseria.ea:efficient.1998s}
\begin{equation}
 (A-B) (A+B) (X+Y) = \omega^2 (X+Y)
\end{equation}
obey the normalization relation;
\begin{equation}
    X_i X_j^{\dagger} - Y_i Y_j^{\dagger} = \delta_{ij} 
\end{equation}
Due to the limitations of converting this rather extensive 
excitation vector $(X+Y)_i$ with dimension 
$N_{\text{occ}} \times N_{\text{virt}}$
into a single picture per excitation, we note that only the most important components are depicted. We cut off the contributions
if $\sum_n max[(X + Y)^2]_n \ge 0.90$ is reached, neglecting 
any remaining contributions. The remaining contributions 
are strongly delocalized. It is noted that this corresponds
to an \textit{ad hoc} conversion of the change in density
described by TD-DFT into a single particle wavefunction 
picture. This approximate method yields a convenient 
reference frame for the spin pairs and unpaired spins 
predicted to be present in the corresponding excited states.

\begin{figure}[h!]
\centering
\includegraphics[width = 0.95\columnwidth]{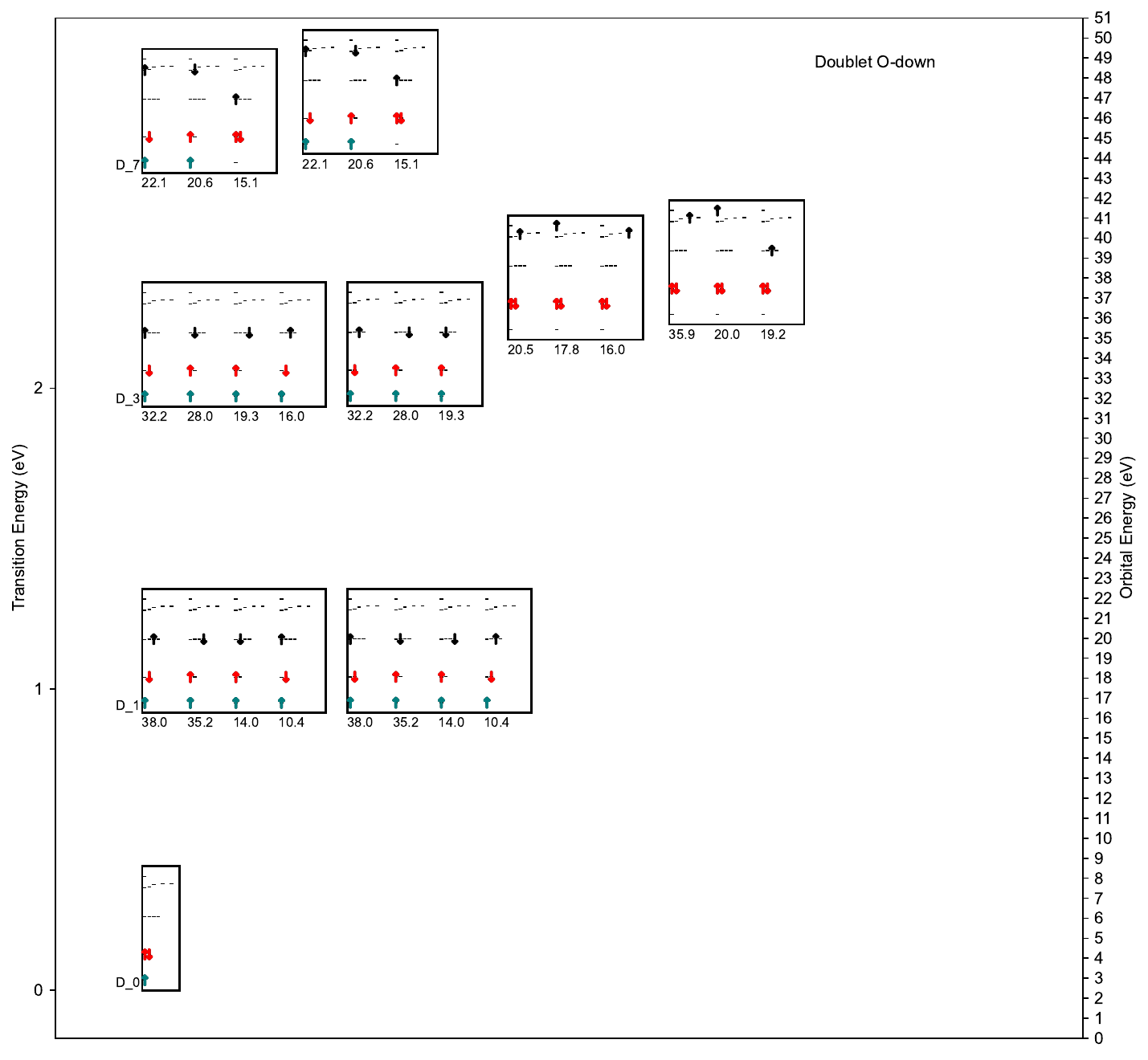}
        \caption{\textbf{Many-body states represented by single particle wavefunctions:} Many-body states for the neutral doublet spin configurations for O-down . For each many-body state, a box comprises the dominant single particle wavefunctions contributing to the different excitations (relative contributions are indicated below the box). Energies of single particle wavefunctions (small horizontal dashes) are plotted on a consistent energy scale (y axis on the right). Energies of many-body states are scaled by a factor of 18 (y axis on the left). Single particle avefunctions 3171beta and 3172alpha are depicted in red, 3171alpha is depicted in teal.} 
\label{fig:SI-Duplett_reverse_ind}
\end{figure}

\begin{figure}[h!]
\centering
\includegraphics[width = 0.95\columnwidth]{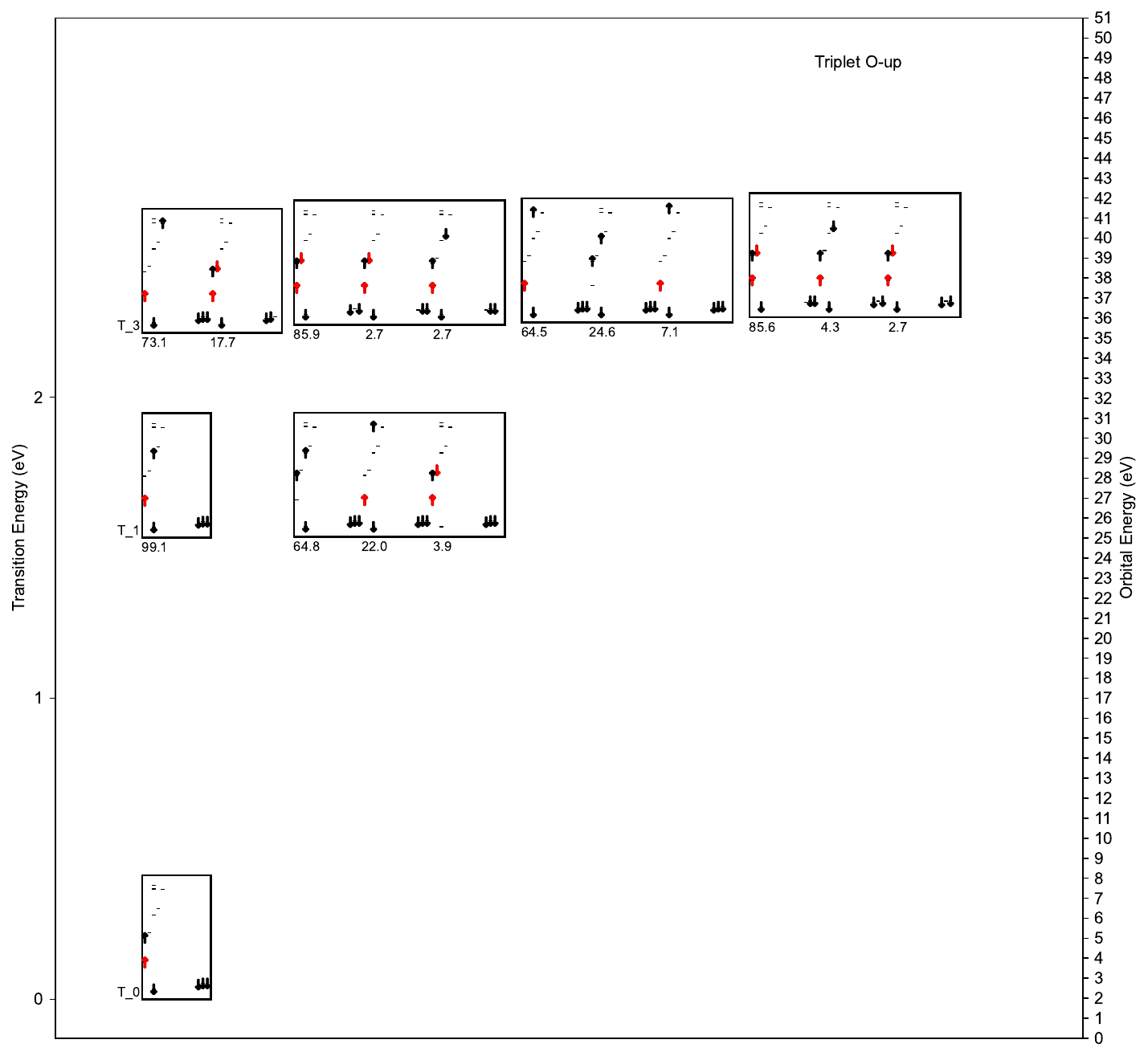}
        \caption{\textbf{Many-body states represented by single particle wavefunctions:} Many-body states for the positively charged VOPc triplet spin configurations for O-up . For each many-body state, a box comprises the dominant single particle wavefunctions contributing to the different excitations (relative contributions are indicated below the box). Energies of single particle wavefunctions (small horizontal dashes) are plotted on a consistent energy scale (y axis on the right). Energies of many-body states are scaled by a factor of 18 (y axis on the left). Single particle avefunctions 3171beta and 3172alpha are depicted in red, 3171alpha is depicted in teal.} 
\label{fig:SI-Triplett_normal_simpl_ind}
\end{figure}

\begin{figure}[h!]
\centering
\includegraphics[width = 0.95\columnwidth]{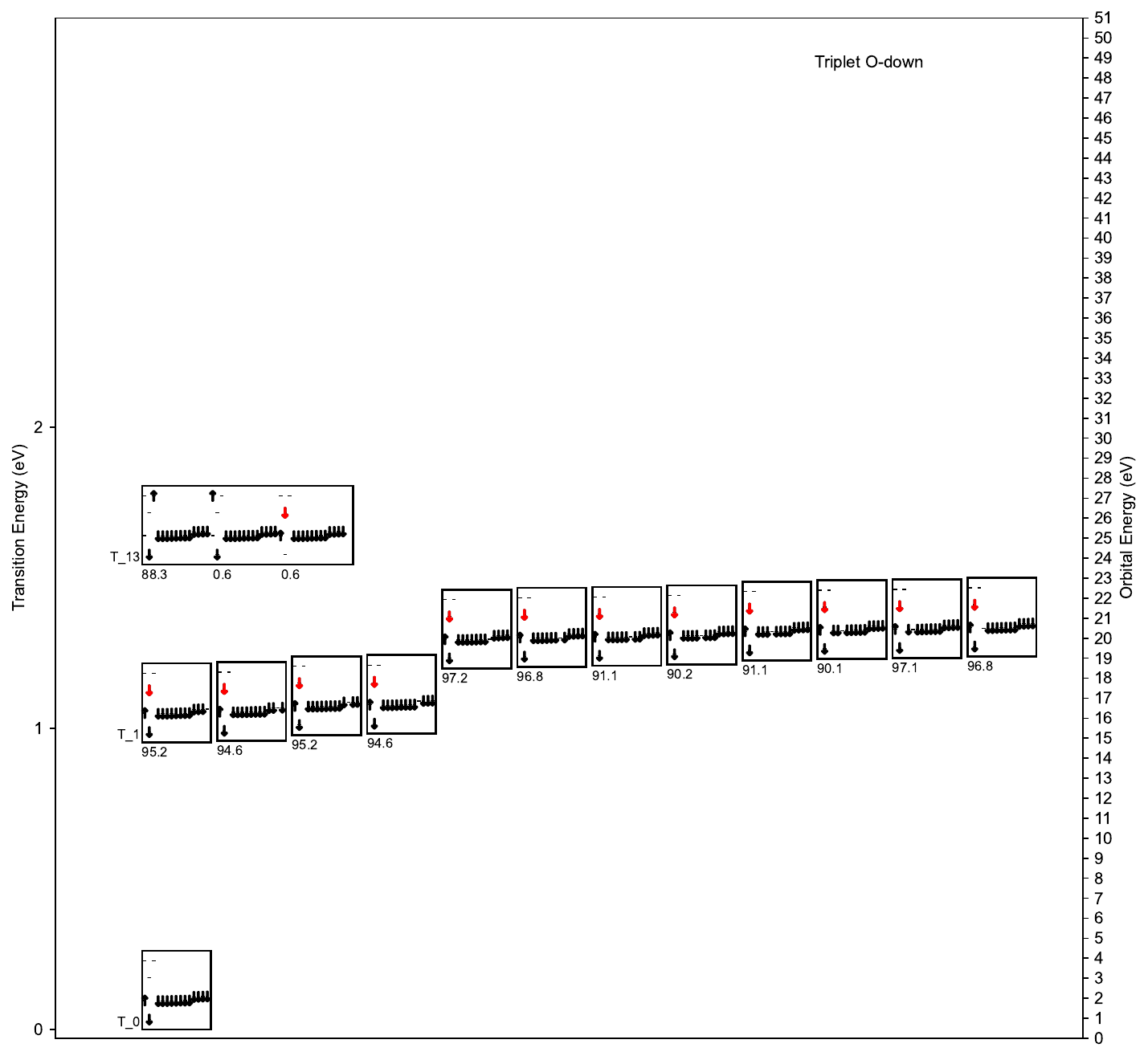}
        \caption{\textbf{Many-body states represented by single particle wavefunctions:} Many-body states for the positively charged VOPc triplet spin configurations for O-down . For each many-body state, a box comprises the dominant single particle wavefunctions contributing to the different excitations (relative contributions are indicated below the box). Energies of single particle wavefunctions (small horizontal dashes) are plotted on a consistent energy scale (y axis on the right). Energies of many-body states are scaled by a factor of 18 (y axis on the left). Single particle avefunctions 3171beta and 3172alpha are depicted in red, 3171alpha is depicted in teal.} 
\label{fig:SI-Triplett_reverse_simpl_ind}
\end{figure}

Specifically for the charged triplet state of the VOPc (O-down) configuration (see S\ref{fig:SI-Triplett_reverse_simpl_ind}), the excitations can be described using a rather simple picture: Each calculated (dark) state T$_1$-T$_{12}$ is an excitation from the HOMO-N to the single occupied (SOMO) orbital, mainly replicating the orbital order. The first bright state, T$_{13}$ is then a complex mixture of several occupied states exciting into LUMO+1/2/3. For the VOPc (O-down) configuration (see S\ref{fig:SI-Triplett_normal_simpl_ind}), instead considerably more involved combinations of orbital products are observed, leading to excitations that are increasingly difficult to describe in a single particle picture. For the uncharged doublet in the O-up configuration (see S\ref{fig:SI-Duplett_normal_ind}), the D$_{1,2}$ excited states can be described as a degenerate
mixture of SOMO to LUMO+1 in both spin channels alpha and beta. The D$_{3,4}$ excited states in the O-up configuration are described in an analogous manner, being a SOMO to LUMO+2 excitation in both spin channels, explaining the degenerate nature. Again, the O-down configuration of VOPc (see S\ref{fig:SI-Duplett_reverse_ind}) is considerably more involved with the D$_{1,2}$ and D$_{3,4}$ excited states being different linear combinations of SOMO to LUMO+1/2, again yielding 2 nearly degenerate energy levels.\\
This is in line with our earlier observations of an increased mixing of single particle orbitals in the VOPc O-down configuration.

\clearpage

\clearpage

\clearpage

\clearpage

\clearpage
\begin{figure}[h!]
\centering
\includegraphics[width = 0.95\columnwidth]{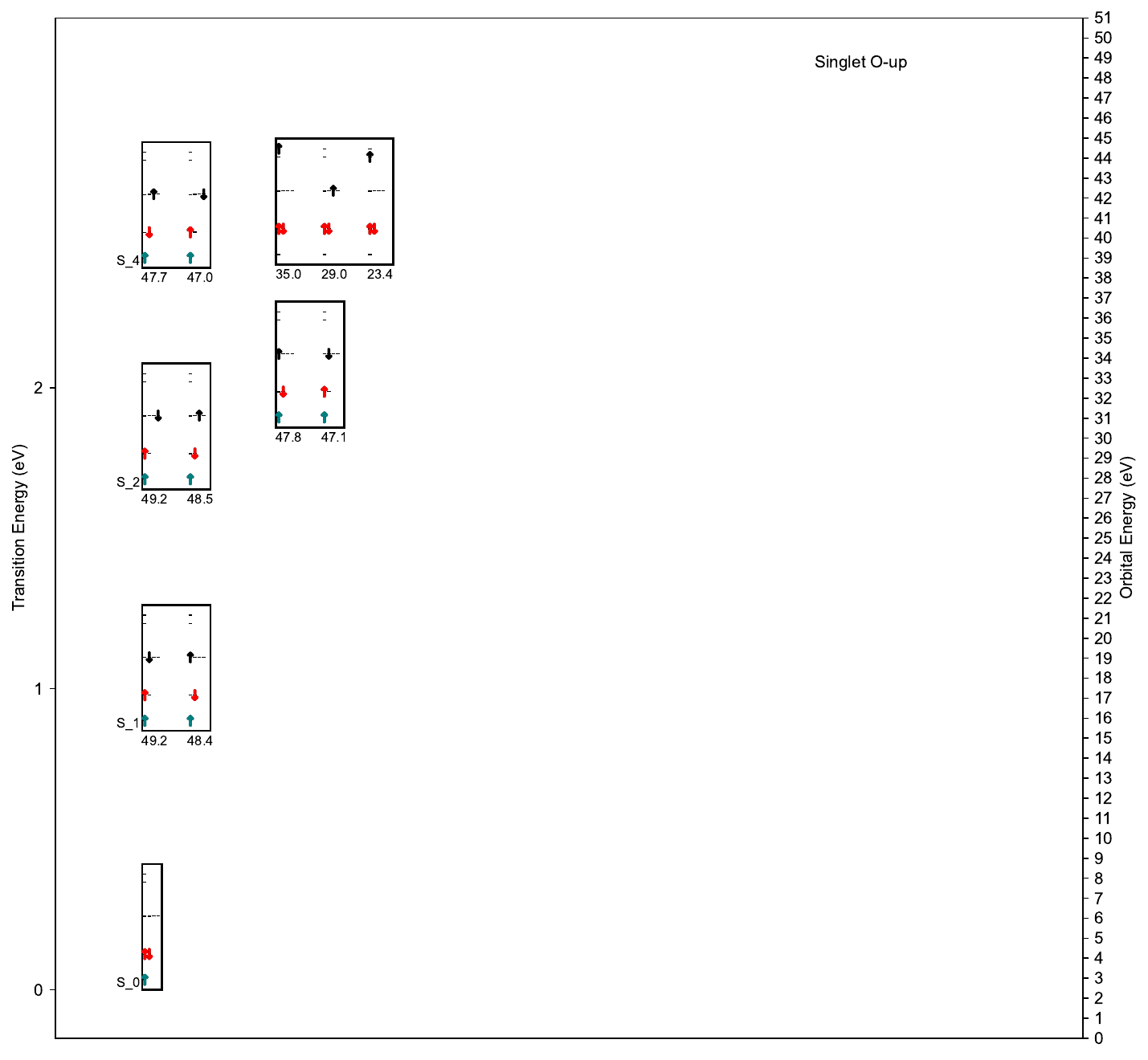}
        \caption{\textbf{Many-body states represented by single particle wavefunctions:} Many-body states for the positively charged VOPc singlet spin configurations for O-up . For each many-body state, a box comprises the dominant single particle wavefunctions contributing to the different excitations (relative contributions are indicated below the box). Energies of single particle wavefunctions (small horizontal dashes) are plotted on a consistent energy scale (y axis on the right). Energies of many-body states are scaled by a factor of 18 (y axis on the left). Single particle avefunctions 3171beta and 3172alpha are depicted in red, 3171alpha is depicted in teal.} 
\label{fig:SI-Singlett_normal_ind}
\end{figure}

\clearpage
\begin{figure}[h!]
\centering
\includegraphics[width = 0.95\columnwidth]{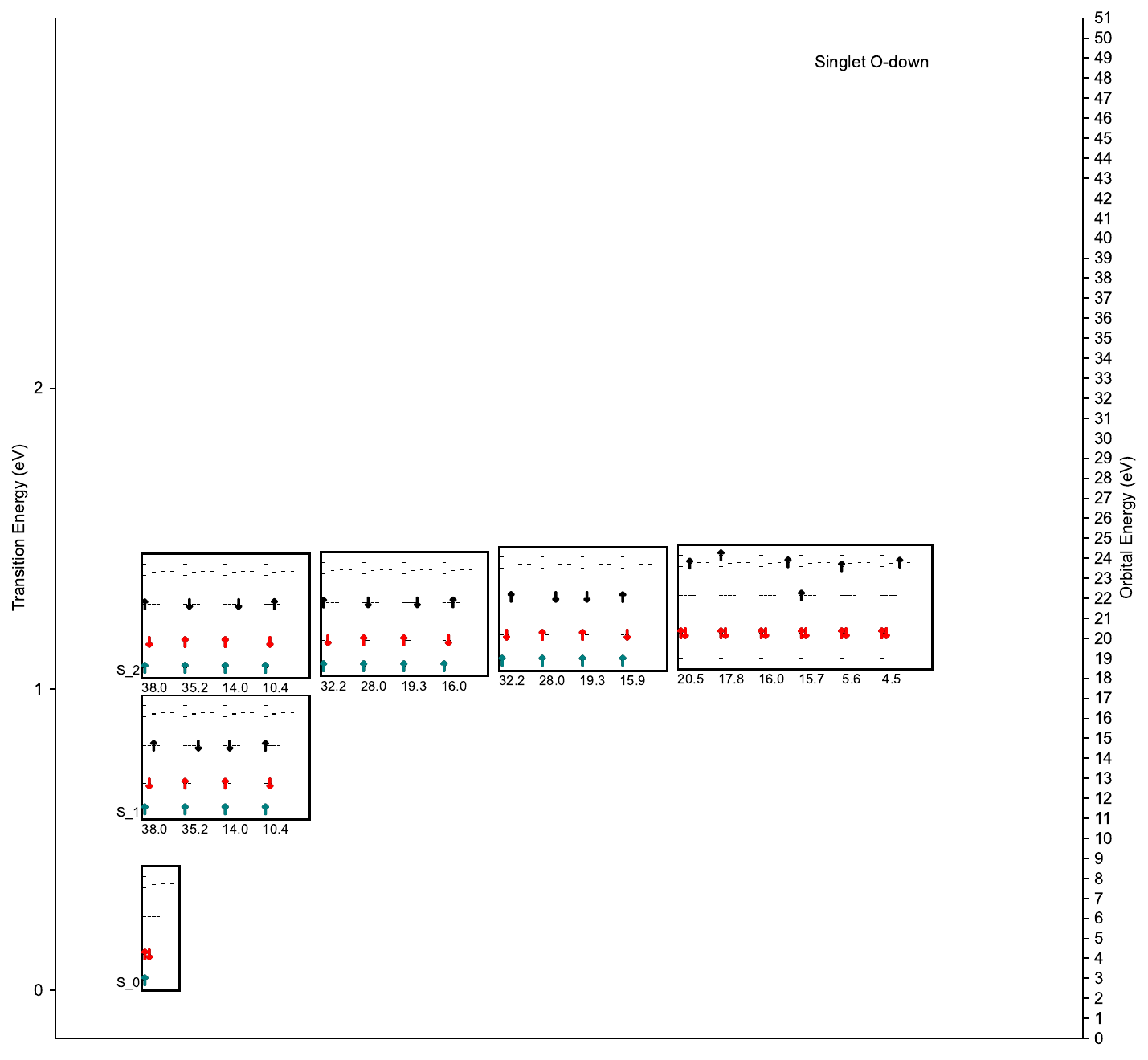}
        \caption{\textbf{Many-body states represented by single particle wavefunctions:} Many-body states for the positively charged VOPc singlet spin configurations for O-down . For each many-body state, a box comprises the dominant single particle wavefunctions contributing to the different excitations (relative contributions are indicated below the box). Energies of single particle wavefunctions (small horizontal dashes) are plotted on a consistent energy scale (y axis on the right). Energies of many-body states are scaled by a factor of 18 (y axis on the left). Single particle avefunctions 3171beta and 3172alpha are depicted in red, 3171alpha is depicted in teal.} 
\label{fig:SI-Singlett_reverse_ind}
\end{figure}

\end{document}